\documentclass[preprint,12pt]{elsarticle}

\usepackage{amsmath,amssymb,graphicx,blkarray}

\DeclareMathOperator{\Normal}{\mathcal{N}}

\journal{Displays}
\title{An empirical-parametric gamma calibration algorithm}

\author[rvt]{Jani Isohanni}
\ead{Jani.Isohanni@tut.fi}
\author[rvt]{Robert Pich\'{e}}
\ead{Robert.Piche@tut.fi}

\address[rvt]{Tampere University of Technology, Department of Automation Science and Engineering, PO Box 692, FI-33720 Tampere, Finland}

\begin{document}



\begin{abstract}
A method to determine the gamma correction curves for displays is presented.
An empirical model is first constructed 
from exhaustive measurements of a few representative units.
The model parameters for the remaining units are 
then fitted using only a few measurements.
The method uses standard least-squares algorithms 
and is computationally light.
Experimental results for a small sample of LCD displays are presented.
\end{abstract}


\maketitle 

\section{Introduction}

Modern display devices generally incorporate a mechanism 
for adjustment of their gamma correction curves.
This mechanism is used in the production line
to correct for variations between units.
In principle, the adjustment could be based on 
an exhaustive spectrographic measurement of each unit,
but this would be time-consuming and expensive.
In fast paced production lines, 
one needs to be able to adjust a device's gamma curve 
with just a few spectrographic measurements.

One such approach is to fit or estimate a few parameters 
of a pre-defined mathematical model of the display,
such as the LCD model of Kwak and MacDonald \cite{Kwak}
or the CTR model of Berns \cite{Berns}.
This approach can provide good results 
if the model can accurately represent the device's gamma curves
with a small number of adjustable parameters.
However, the gamma curves of
a given production batch may have some peculiar characteristics
that cannot easily be represented
by a standard general-purpose model,
and one then faces the problem of modifying the model
to remove systemic deviations in the fitted curves.
Existing mathematical models may also need to be modified
when new display technologies or manufacturing methods are introduced.

In this work,
we propose a  gamma calibration method 
that combines the accuracy of exhaustive measurement
with the efficiency of a parametric model.
First,
a parametric model is constructed 
from exhaustive measurements of a few representative units.
Then, 
for each unit in the production line,
the parameters of the model are fitted using only a few measurements.
Because the method uses only standard least-squares algorithms,
it is computationally light and reliable.

\section{Preliminaries}

In our setting the gamma response of a display unit is measured from grey levels.
The unit is 
given 
an sRGB input with a grey color from white $(0,0,0)$ to black $(255,255,255)$,
and the color that is displayed without any gamma correction is measured.
The measurements are 
transformed to sRGB color space and scaled to the range $(0,0,0)-(255,255,255)$.
 
The actual 
gamma correction  is generally applied to each RGB channel separately,
 thus
we will study the channels separately, and the resulting algorithm is to be 
applied to each channel accordingly. The separation of color channels 
and the assumption of
channel independence is a usual characteristic of display control systems, as mentioned by 
Kazuhiro \cite{kazuhiro}.

Now, we have $J$ display units measured thoroughly with responses $d_{i,j}$, where
subindex $i$ indicates the measured grey level and $j$ the display unit. 
We call these $J$ units the \emph{training sample}.
The measurements are done on $I$ different grey levels $x_i\in (0...255)$.

For the unit that we are to calibrate, we have $P$ responses $d'_p$ at grey levels
$x_{i_p}$, where we have $P\leq J$ and $p\in[1...P]$, $i_p\in I$, where $I$ is an 
index set with the indices of the measured colors.
Our goal is to estimate
the responses at other grey levels (the gamma correction curve)
accurately even when $P$ is small. 

\section{The method}

Let $A$ be the $J$-by-$256$ matrix of the training set responses of a color channel,
say the red channel. An element $a_{i,j}$ of $A$ is the response to
input $(j-1)$th input grey level of 
the $i$th display.

We estimate the response of this color channel of the test unit one 
grey level at a time, as follows.
Suppose we are estimating the response for input level $x$.
We first extract the columns of $A$ that correspond to the measured
values and the columns that correspond to the input level  value we are to estimate. 
Thus we get a $J$-by-$(P+1)$ matrix $A_x$.
For convenience we put values corresponding to $x$ in the first column.

The rows of $A_x$ are now treated as samples from $(P+1)$-dimensional
normal distribution 
with mean $\mu$ and covariance $\Sigma$.
Both $\mu$ and $\Sigma$ can be estimated from the rows of $A_x$ 
 by standard techniques. 

Consider now the measurements we have from the test unit. We have $P$
measured values and one unknown. We arrange these to a vector
$(x,\vec{y})$ where $\vec{y}$ models the measured values and $x$ is unknown.
Notice that the order of elements of $(x,\vec{y})$ naturally follows that of the matrix $A_x$.
The vector $(x,\vec{y})$ can now be thought of as a sample from the distribution
$\Normal_x(\mu,\Sigma)$ and we would like to estimate how $x$ is distributed
when we know the value of $\vec{y}$.

Since the value of $\vec{y}$ is known we can estimate the distribution of 
$x$ with a conditional distribution.
The conditional distribution of one of the dimensions is constructed as follows. 
First we partition $\mu$ 
 as 
 \begin{equation*}
\mu = \begin{blockarray}{ccc}
 & {\scriptstyle 1} & {\scriptstyle P}\\
\begin{block}{c(cc)}
{\scriptstyle 1} & \mu_1 & \mu_2 \\
\end{block}
\end{blockarray}
\end{equation*} 
and $\Sigma$ as
\begin{equation}
\label{covariance}
\Sigma = \begin{blockarray}{ccc} 
 {\scriptstyle 1} & {\scriptstyle P} & \, \\
 \begin{block}{(cc)c}
\Sigma_{11} & \Sigma_{12}  & \,\,\,\, {\scriptstyle 1} \\
\Sigma_{21} & \Sigma_{22}  & \,\,\,\, {\scriptstyle P} \\
\end{block}
\end{blockarray}
\end{equation}
We are now to find the conditional distribution of $x$ given $\vec{y}=a$, i.e.
$(x\mid\vec{y}=a)$, where $a$
contains the values measured from the display for which we want the estimate.
The conditional distribution of $(x\mid\vec{y}=a)$ is
$\Normal(\mu',\Sigma')$ where
\begin{equation*}
\mu' = \mu_1 + \Sigma_{12}\Sigma_{22}^{-1}(a-\mu_2)
\end{equation*}
and
\begin{equation*}
\Sigma' = \Sigma_{11} - \Sigma_{12}\Sigma_{22}^{-1}\Sigma_{21}.
\end{equation*}
Here the matrix $\Sigma_{22}^{-1}$ is generalized inverse, also known Moore-Penrose inverse,
 of $\Sigma_{22}$, see \cite{Demmel} for further details.
Now the predicted value for $x$ is $\mu'$ and $\Sigma'$ provides us confidence
intervals if wanted.

This algorithms computes the 
 pseudoinverse $\Sigma_{22}^{-1}$ via singular value
decomposition because it provides good numerical stability. For further information
see for example Golub \& Loan \cite{golub2012matrix}.

\section{Experiments}

\begin{figure}[htbp]
\centerline{\includegraphics[width=0.6\columnwidth]{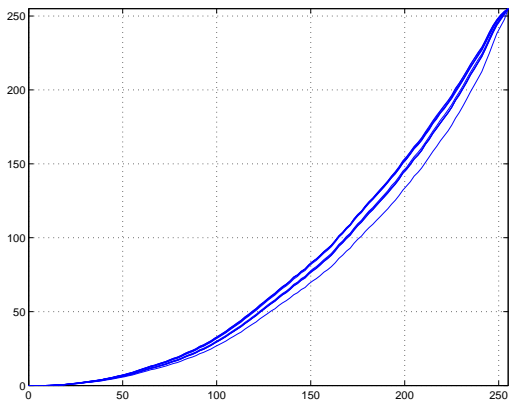}}
\caption{Responses of training sample of red channel.}
\label{all}
\end{figure}

We tested the method with a sample of ten displays.
The displays in the sample were measured on blackÊ$(0,0,0)$ and grey values  $(1,1,1)$,
$(3,3,3)$,..., $(255,255,255)$ totaling to $129$ measured grey levels.
The results for red channel are shown in Figure \ref{all}.
We notice the slightly clustered nature of the displays and one display that variates from
the others. The used scaling guarantees us that on $(0,0,0)$ and $(255,255,255)$ all responses
are equal.

We treated nine of the displays
as a training sample and predicted the response values for the remaining unit. 
The tests were run on
red channel.

\begin{figure*}[htbp]
\label{virheet123}
\centerline{\includegraphics[width=1\columnwidth]{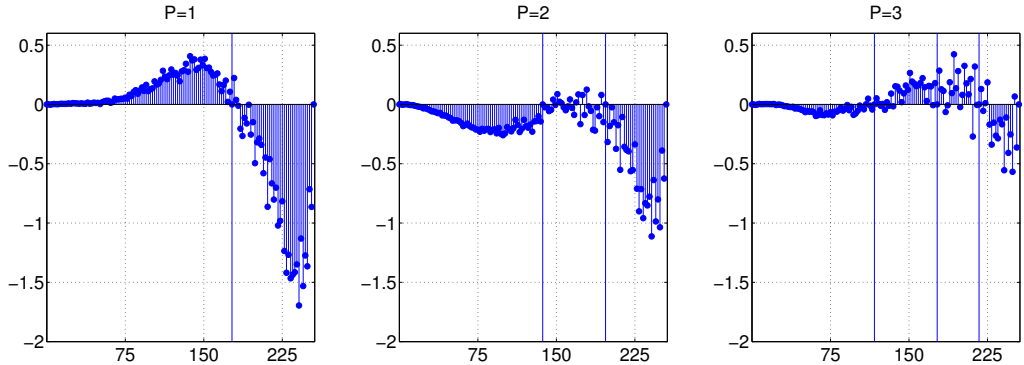}}
\caption{The error in red color channel with one, two and three measured grey levels.
Vertical lines indicate the measured valued.}
\end{figure*}

We tested the algorithm with 
one, two and three measured grey levels, thus $P$ is
$1$, $2$, and $3$ respectively. The tests were run on various displays and the results
are practically uniform across the data.
The results for the chosen test unit on red channel 
are in Figure \ref{virheet123}.
Notice
that the error escalates slightly when we drop the
number of measurements to two and even more when only one value is used on the prediction.
It is readily seen that with only three grey levels we reach quite satisfying 
prediction. One should notice that four grey levels would make the fit worse in the test case
because we have only nine displays in the training sample, which makes the
estimate of the covariance matrix problematic. 

\begin{figure*}[htbp]
\label{virheet}
\centerline{\includegraphics[width=1\columnwidth]{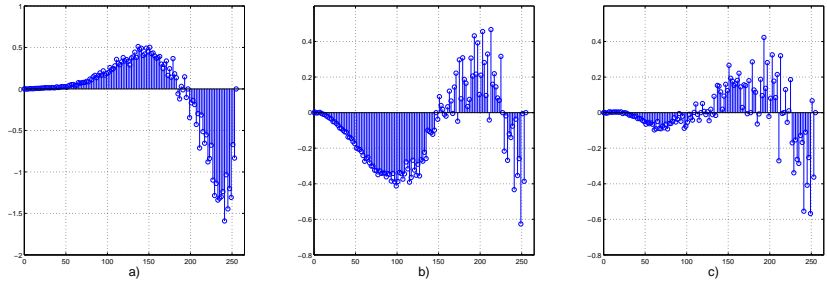}}
\caption{Error in prediction for the unit by color channel. }
\end{figure*}

For the test unit the prediction error is of the same magnitude for each color channel.
Figure \ref{virheet} shows the prediction errors for each color channel when
three grey levels were measured: $(117,117,117)$, $(177,177,177)$ and $(217,217,127)$.
Unfortunately there is no established method for choosing the grey levels. The values
were first chosen to be almost equispaced and later finetuned to the chosen values. The 
differences between tested values were not big.

\section{Conclusions}

We used the assumption that the data is normal distributed. Figure \ref{all} does,
however, suggest that a suitable bimodal distribution might provide better results.
In any case, the data set is relatively small and not much can be inferred 
about the (assumed) underlying distribution.

Also, a larger training sample would provide more freedom on choosing the number of
measurements. Now the number of measurements we can use is limited by the amount
of data, because estimating the covariance matrix becomes problematic. In theory
one could overcome the situation with a least-square fit but in our tests the
least-square solution proved to introduce big errors on bright values.

\section*{Acknowledgements}

The research was supported by Nokia corporation. The authors would like to
express their gratitude to
Mika Antila and Jussi Ropo for introducing the problem and providing
the test data.




\end{document}